\documentclass[a4paper,10pt]{article}

\usepackage{epsfig}

\newcommand{\CTT}{C^{\rm TT}_\ell}
\newcommand{\CTE}{C^{\rm TE}_\ell}
\newcommand{\CEE}{C^{\rm EE}_\ell}
\newcommand{\CBB}{C^{\rm BB}_\ell}
\newcommand{\CTB}{C^{\rm TB}_\ell}
\newcommand{\CEB}{C^{\rm EB}_\ell}
\newcommand{\EI}{{\cal E}_{\rm Inf}}

\def\ensuremath#1{$#1$}

\newcommand{\BOOMERANG}{Boom2K flight}

\newcommand{\mpc}{{\rm\,Mpc}}

\def\lsim{\mathrel{\rlap{\lower4pt\hbox{\hskip1pt$\sim$}}
    \raise1pt\hbox{$<$}}}                
\def\gsim{\mathrel{\rlap{\lower4pt\hbox{\hskip1pt$\sim$}}
    \raise1pt\hbox{$>$}}}                

\begin{document}
\markboth{Cosmology with CMB anisotropy}{Tarun Souradeep}

\title{Cosmology with CMB anisotropy}

\author{Tarun Souradeep\\Inter-University Centre for Astronomy and
Astrophysics,\\ Post Bag 4, Ganeshkhind, Pune 411~007, India.\\
tarun@iucaa.ernet.in}
\date{}
\maketitle

\footnotetext[1]{Invited plenary talk at the IXth. International
Workshop on High Energy Physics Phenomenology (WHEPP-9), Institute of
Physics, Bhubaneshwar, Jan 3-14, 2006.}
\begin{abstract}
Measurements of CMB anisotropy and, more recently, polarization have
played a very important role allowing precise determination of various
parameters of the `standard' cosmological model.  The expectation of
the paradigm of inflation and the generic prediction of the simplest
realization of inflationary scenario in the early universe have also
been established -- `acausally' correlated initial perturbations in a
flat, statistically isotropic universe, adiabatic nature of primordial
density perturbations. Direct evidence for gravitational instability
mechanism for structure formation from primordial perturbations has
been established. In the next decade, future experiments promise to
strengthen these deductions and uncover the remaining crucial
signature of inflation -- the primordial gravitational wave
background.  
\end{abstract}

\section{Introduction}	
 

The transition to precision cosmology has been spearheaded by
measurements of CMB anisotropy and, more recently, polarization.  Our
understanding of cosmology and structure formation necessarily depends
on the relatively unexplored physics of the early universe that
provides the stage for scenarios of inflation (or related
alternatives).  The CMB anisotropy and polarization contains
information about the hypothesized nature of random primordial/initial
metric perturbations -- (Gaussian) statistics, (nearly scale
invariant) power spectrum, (largely) adiabatic vs.  iso-curvature and
(largely) scalar vs. tensor component.  The `default' settings in
bracket are motivated by inflation.  Estimation of cosmological
parameters implicitly depend on the assumed values of the initial
conditions, or, explicitly on the scenario of generation of initial
perturbations~\cite{recons_us}.  Besides precise determination of
various parameters of the `standard' cosmological model, observations
have also established some important basic tenets of cosmology and
structure formation in the universe -- `acausally' correlated initial
perturbations, adiabatic nature primordial density perturbations,
gravitational instability as the mechanism for structure formation. We
have inferred a spatially flat universe where structures form by the
gravitational evolution of nearly scale invariant, adiabatic
perturbations in a predominant form of non--baryonic cold dark matter
which is sub-dominant to a form dark energy that does not cluster (on
astrophysical scales).

The signature of primordial perturbations on super-horizon scales at
decoupling in the CMB anisotropy and polarization are the most
definite evidence for new physics (eg., inflation ) in the early
universe that needs to be uncovered.  We briefly review the
observables from the CMB sky and importance to understanding cosmology
in section~\ref{cmb} The article briefly summarizes the recent
estimates of the cosmological parameters and highlight the success of
recent cosmological observations in establishing some of the
fundamental tenets of cosmology and structure~:

\begin{itemize}

\item{} Primordial perturbations from Inflation.(Sec.~\ref{PI});

\item{} Gravitational instability mechanism for structure
formation(Sec.~\ref{GI});

\item{} Statistical Isotropy of the universe (Sec.~\ref{SI}).

\end{itemize}

At this time, the attention of the community is largely focused on
estimating the cosmological parameters.  The next decade would see
increasing efforts to observationally test fundamental tenets of the
cosmological model using the CMB anisotropy and polarization
measurements (and related LSS observations, galaxy survey,
gravitational lensing, etc.).

\section{CMB observations and cosmological parameters}
\label{cmb}

The angular power spectra of the Cosmic Microwave Background
temperature fluctuations ($C_\ell$)have become invaluable observables
for constraining cosmological models. The position and amplitude of
the peaks and dips of the $C_\ell$ are sensitive to important
cosmological parameters, such as, the relative density of matter,
$\Omega_0$; cosmological constant, $\Omega_\Lambda$; baryon content,
$\Omega_B$; Hubble constant, $H_0$ and deviation from flatness
(curvature), $\Omega_K$.

\begin{figure}
\begin{center}
\includegraphics[scale= 0.35,angle=-90]{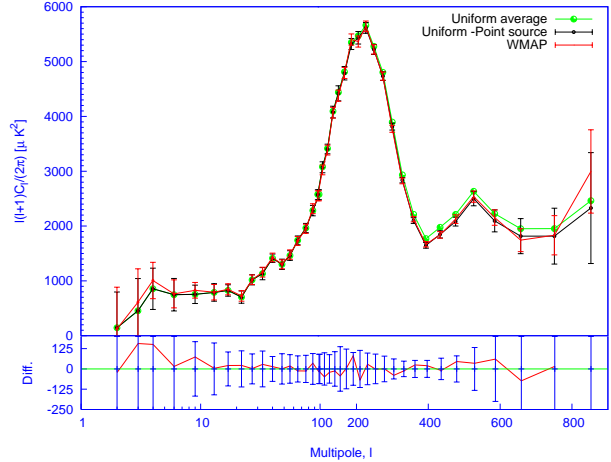}
\caption{The angular power spectrum estimated from the
WMAP multi-frequency using a self-contained model free approach to
foreground removal \protect{\cite{sah06}}
(black curve) is compared to
the WMAP team estimate (red). The published binned WMAP power
spectrum plotted in red line with error bars for comparison. The lower
panel shows the difference in the estimated power spectra. The method
holds great promise for CMB polarization where modeling uncertainties
for foregrounds are much higher.}
\label{WMAPCL}
\end{center}
\end{figure}

  The angular spectrum of CMB temperature fluctuations has been
measured with high precision on large angular scales ($\ell < 800$) by
the WMAP experiment \cite{hin_wmap06}, while smaller angular scales
have been probed by ground and balloon-based CMB experiments
\cite{ruh02,read04,dick04,kuo04,hal02}.  These data are broadly
consistent with a $\Lambda$CDM model in which the Universe is
spatially flat and is composed of radiation, baryons, neutrinos and,
the exotic, cold dark matter and dark energy. The exquisite
measurements by the Wilkinson Microwave Anisotropy Probe (WMAP) mark a
successful decade of exciting CMB anisotropy measurements and are
considered a milestone because they combine high angular resolution
with full sky coverage and extremely stable ambient condition (that
control systematics) allowed by a space mission .  Figure~\ref{WMAPCL}
shows the angular power spectrum of CMB temperature fluctuations
obtained from the first year of WMAP data~\cite{sah06}.  The third
year of WMAP observations have also included CMB polarization
results. The WMAP results are of excellent quality and show robustness
to different analysis methods~\cite{wmap3reanal}.

One of the firm predictions of this working `standard' cosmological
model is linear polarization pattern ($Q$ and $U$ Stokes parameters)
imprinted on the CMB at last scattering surface. Thomson scattering
generates CMB polarization anisotropy at decoupling~\cite{cmb_polar}.
A net pattern of linear polarization is retained due to local
quadrupole intensity anisotropy of the CMB radiation impinging on the
electrons at $z_{rec}$. The coordinate--free description decomposes
the two kinds of polarization pattern on the sky based on their
different parities.  In the spinor approach, the even parity pattern
is called the $E$--mode and the odd parity pattern the $B$--mode.
With the introduction of polarization, there are a total of 4 power
spectra to determine: $\CTT, \CTE, \CEE, \CBB$. Parity conservation
~\footnote{On the other hand, a non-zero detection of $\CTB$ or
$\CEB$, over and above observational artifacts, could be tell-tale
signatures of exotic parity violating physics~\cite{parviol}.}
eliminates the two other possible power spectra, $\CTB$ \&
$\CEB$. While CMB temperature anisotropy can also be generated during
the propagation of the radiation from the last scattering surface, the
CMB polarization signal can be generated only at the last scattering
surface, where the optical depth transits from large to small
values. The polarization information complements the CMB temperature
anisotropy by isolating the effect at the last scattering surface from
effects along the line of sight.


The CMB polarization is an even cleaner probe of early universe
scenarios, that promises to complement the remarkable successes of CMB
anisotropy measurements. 
The CMB
polarization signal is much smaller than the anisotropy
signal. Measurements of polarization at sensitivities of $\mu K$
(E-mode) to tens of $nK$ level (B-mode) pose spectacular challenges
for ongoing and future experiments.

 After the first detection of CMB polarization by DASI in 2003, the
field has rapidly grown, with measurements coming in from a host of
ground--based and balloon--borne dedicated CMB polarization
experiments. The Degree Angular Scale Interferometer (DASI) measured
the CMB polarization spectrum over a limited band of angular scales
($l\sim 200-440$) in late 2002~\cite{kov_dasi02}.  The DASI experiment
recently published results of much refined measurements with 3 years
of data~\cite{dasi_3y}. More recently, the Boomerang collaboration
reports measurements of $\CTT$, $\CTE$ and $\CEE$ and a non--detection
of $B$--modes~\cite{boom_polar}. The recent release of full sky E-mode
polarization maps and polarization spectra by WMAP are a new milestone
in CMB research~\cite{pag_wmap06,kog_wmap03}.  As expected, there has
been no detection of cosmological signal in B-mode of
polarization. The lack of $B$--mode power suggests that foreground
contamination is at a manageable level which is good news for future
measurements.  Scheduled for launch in 2007, the Planck satellite will
greatly advance our knowledge of CMB polarization by providing
foreground/cosmic variance--limited measurements of $\CTE$ and $\CEE$
out beyond $l\sim 1000$.  We also expect to detect the lensing signal,
although with relatively low precision, and could see gravity waves at
a level of $r\sim 0.1$. In the future, a dedicated CMB polarization
mission has been listed as a priority by both NASA (Beyond Einstein)
and ESA (Cosmic Vision) in the time frame 2015-2020.  These primarily
target the $B$-mode polarization signature of gravity waves, and
consequently, identify the viable sectors in the space of inflationary
parameters.
\begin{table}[h] 
\begin{center}
\caption{The table taken from 
~Ref.\protect{\cite{sper_wmap06}}
summarizes the estimated values of the cosmological parameters of the
$\Lambda$CDM Model. The best fit parameters correspond to the maximum
of the joint likelihoods of various combinations of CMB anisotropy and
large scale structure data.}  {\begin{tabular}{|c||c|c|c|c|} 
\hline &&&& \\
{Data combo.$\rightarrow$} &WMAP & WMAP& WMAP+ACBAR & WMAP + \\ &Only & +CBI+VSA &
+BOOMERanG &2dFGRS \\ {Parameters $\downarrow$} & & & & \\ \hline
\hline
 &&&& \\
100$\Omega_b h^2$ & 
\ensuremath{2.233^{+ 0.072}_{- 0.091} \mbox{ }} &
\ensuremath{2.212^{+ 0.066}_{- 0.084} \mbox{ }} &
\ensuremath{2.231^{+ 0.070}_{- 0.088} \mbox{ }} &
\ensuremath{2.223^{+ 0.066}_{- 0.083} \mbox{ }}  \\
$\Omega_m h^2 $ & 
\ensuremath{0.1268^{+ 0.0072}_{- 0.0095} \mbox{ }} &
\ensuremath{0.1233^{+ 0.0070}_{- 0.0086} \mbox{ }} &
\ensuremath{0.1259^{+ 0.0077}_{- 0.0095} \mbox{ }} &
\ensuremath{0.1262^{+ 0.0045}_{- 0.0062} \mbox{ }}  \\
$h$ & 
\ensuremath{0.734^{+ 0.028}_{- 0.038} \mbox{ }} &
\ensuremath{0.743^{+ 0.027}_{- 0.037} \mbox{ }} &
\ensuremath{0.739^{+ 0.028}_{- 0.038} \mbox{ }} &
\ensuremath{0.732^{+ 0.018}_{- 0.025} \mbox{ }}  \\
$A$ & 
\ensuremath{0.801^{+ 0.043}_{- 0.054} \mbox{ }} &
\ensuremath{0.796^{+ 0.042}_{- 0.052} \mbox{ }} &
\ensuremath{0.798^{+ 0.046}_{- 0.054} \mbox{ }} &
\ensuremath{0.799^{+ 0.042}_{- 0.051} \mbox{ }}  \\
$\tau$ & 
\ensuremath{0.088^{+ 0.028}_{- 0.034} \mbox{ }} &
\ensuremath{0.088^{+ 0.027}_{- 0.033} \mbox{ }} &
\ensuremath{0.088^{+ 0.030}_{- 0.033} \mbox{ }} &
\ensuremath{0.083^{+ 0.027}_{- 0.031} \mbox{ }}  \\
$n_s$ & 
\ensuremath{0.951^{+ 0.015}_{- 0.019} \mbox{ }} &
\ensuremath{0.947^{+ 0.014}_{- 0.017} \mbox{ }} &
\ensuremath{0.951^{+ 0.015}_{- 0.020} \mbox{ }} &
\ensuremath{0.948^{+ 0.014}_{- 0.018} \mbox{ }}  \\
&&&& \\\hline 
\hline &&&& \\
$\sigma_8$ & 
\ensuremath{0.744^{+ 0.050}_{- 0.060} \mbox{ }} &
\ensuremath{0.722^{+ 0.043}_{- 0.053} \mbox{ }} &
\ensuremath{0.739^{+ 0.047}_{- 0.059} \mbox{ }} &
\ensuremath{0.737^{+ 0.033}_{- 0.045} \mbox{ }} 
\\
$\Omega_m $ & 
\ensuremath{0.238^{+ 0.030}_{- 0.041} \mbox{ }} &
\ensuremath{0.226^{+ 0.026}_{- 0.036} \mbox{ }} &
\ensuremath{0.233^{+ 0.029}_{- 0.041} \mbox{ }} &
\ensuremath{0.236^{+ 0.016}_{- 0.024} \mbox{ }} \\
 &&&& \\
\hline
\end{tabular}\label{tab:lcdm_low}}
\end{center}
\end{table}

The measurements of the anisotropy in the cosmic microwave background
(CMB) over the past decade has led to `precision cosmology'.
Observations of the large scale structure in the distribution of
galaxies, high redshift supernova, and more recently, CMB
polarization, have provided the required complementary
information. The current up to date status of cosmological parameter
estimates from joint analysis of CMB anisotropy and Large scale
structure (LSS) data is usually best to look up in the parameter
estimation paper accompanying the most recent results announcement of
a major experiment, such as recent WMAP release~\cite{sper_wmap06}.
Using WMAP data only, the best fit values for cosmological parameters
for the power-law, flat, $\Lambda$CDM model are $(\Omega_m h^2,
\Omega_b h^2, h, n_s, \tau, \sigma_8) =$ $(0.127^{+0.007}_{-0.013},
0.0223^{+0.0007}_{-0.0009}$, $0.73^{+0.03}_{-0.03}$,
$0.951^{+0.015}_{-0.019}$, $0.09^{+0.03}_{-0.03}$,
$0.74_{-0.06}^{+0.05})$.  Table~\ref{tab:lcdm_low} summarizes best fit
parameters that correspond to the maximum of the joint likelihoods (in
a multi-dimensional parameter space) of various combinations of CMB
anisotropy and large scale structure data.

\section{Primordial perturbations from Inflation}
\label{PI}

Any observational comparison based on the structure formation in the
universe necessarily depends on the assumed initial conditions
describing the primordial seed perturbations.  It is well appreciated
that in `classical' big bang model the initial perturbations would
have had to be generated `acausally'. Besides resolving a number of
other problems of classical Big Bang, inflation provides a mechanism
for generating these apparently `acausally' correlated primordial
perturbations~\cite{inflpert}.

 The power in the CMB temperature anisotropy at low multipoles
($l\lsim 60$) first measured by the COBE-DMR~\cite{cobedmr} did
indicate the existence of correlated cosmological perturbations on
super Hubble-radius scales at the epoch of last scattering, except for
the (rather unlikely) possibility of all the power arising from the
integrated Sachs-Wolfe effect along the line of sight. Since the
polarization anisotropy is generated only at the last scattering
surface, the negative trough in the $C_l^{TE}$ spectrum at $l\sim 130$
(that corresponds to a scale larger than the horizon at the epoch of
last scattering) measured by WMAP first sealed this loophole, and
provides an unambiguous proof of apparently `acausal' correlations in
the cosmological perturbations~\cite{pag_wmap06,kog_wmap03,ben_wmap03}. 

Besides, the entirely theoretical motivation of the paradigm of
inflation, the assumption of Gaussian, random adiabatic scalar
perturbations with a nearly scale invariant power spectrum is arguably
also the simplest possible choice for the initial perturbations.  What
has been truly remarkable is the extent to which recent cosmological
observations have been consistent with and, in certain cases, even
vindicated the simplest set of assumptions for the initial conditions
for the (perturbed) universe discussed below.

\subsection{Nearly zero curvature of space}

The most interesting and robust constraint obtained in our quests in
the CMB sky is that on the spatial curvature of the universe. The
combination of CMB anisotropy, LSS and other observations can pin down
the universe to be flat, $\Omega_K \approx-0.02\pm 0.02$. This is
based on the basic geometrical fact that angular scale subtended in
the sky by the acoustic horizon would be different in a universe with
uniform positive (spherical), negative (hyperbolic), or, zero
(Euclidean) spatial curvature.  Inflation dilutes the curvature of the
universe to negligible values and generically predicts a (nearly)
Euclidean spatial section.

The CMB data~\cite{boom_polar} alone places a constraint on the
curvature which is $\Omega_k = -0.037^{+0.033}_{-0.039}$.  Addition of
the LSS data, yields a median value of $\Omega_k = -0.027 \pm 0.016$.
Restricting $H_0$ by the application of a Gaussian HST prior, the
curvature density determined from the \BOOMERANG\ data set and all
previous CMB results was $\Omega_k = -0.015 \pm 0.016$.  A constraint
$\Omega_k =-0.010\pm 0.009$ obtained by combining CMB data with the
red luminous galaxy clustering data, which has its own signature of
baryon acoustic oscillations \cite{eis_sdss05}. The WMAP 3 year data
can (jointly) constrain $\Omega_k =-0.024^{+0.016}_{-0.013}$ even when
allowing for dark energy with arbitrary (constant) equation state
$w$~\cite{sper_wmap06}. (The corresponding joint limit from WMAP-3yr
on the equation of state is also impressive,
$w=-1.062^{+0.128}_{-0.079}$).

\subsection{Adiabatic primordial perturbation}

The polarization measurements provides an important test on the
adiabatic nature primordial scaler fluctuations~\footnote{ Another
independent observable is the baryon oscillation in LSS discussed in
sec~\ref{GI}}.  CMB polarization is sourced by the anisotropy of the
CMB at recombination, $z_{rec}$, the angular power spectra of
temperature and polarization are closely linked.  Peaks in the
polarization spectra are sourced by the velocity term in the same
acoustic oscillations of the baryon-photon fluid at last
scattering. Hence, a clear indication of the adiabatic initial
conditions is the compression and rarefaction peaks in the temperature
anisotropy spectrum be `out of phase' with the gradient (velocity)
driven peaks in the polarization spectra.

The recent measurements of the angular power spectrum the E-mode of
CMB polarization at large $l$ from experiments such as Boomerang2K,
DASI, CAPMAP and CBI have confirmed that the peaks in the spectra are
out of phase with that of the temperature anisotropy spectrum.  Data
from other are comparable. The data is good enough to indicate that
the peaks in EE and TE are out of phase with that of TT as expected
for adiabatic initial conditions~\cite{boom_polar}.  These conclusions
are further borne out in the recent polarization results from the
three years of WMAP data~\cite{pag_wmap06}.


\subsection{Nearly scale-invariant power spectrum ?}
 
In a simple power law parametrization of the primordial spectrum of
density perturbation ($|\delta_k|^2 = A k^{n_s}$), the scale invariant
spectrum corresponds to $n_s=1$. Recent estimation of (smooth)
deviations from scale invariance favor a nearly scale invariant
spectrum~\cite{sel04}.  

Many model-independent searches have also been made to look for
features in the CMB power
spectrum~\cite{bridle03,hanne04,pia03,pia05}.  Accurate measurements
of the angular power spectrum over a wide range of multipoles from the
WMAP has opened up the possibility to deconvolve the primordial power
spectrum for a given set of cosmological
parameters~\cite{max_zal02,mat_sas0203,shaf_sour04,bump05}.  The
primordial power spectrum has been deconvolved from the angular power
spectrum of CMB anisotropy measured by WMAP using an improved
implementation of the Richardson-Lucy algorithm~\cite{shaf_sour04}.
The most prominent feature of the recovered primordial power spectrum
is a sharp, infra-red cut off on the horizon scale. It also has a
localized excess just above the cut-off which leads to great
improvement of likelihood over the simple monotonic forms of model
infra-red cut-off spectra considered in the post WMAP literature.  The
form of infra-red cut-off is robust to small changes in cosmological
parameters.  Remarkably similar form of infra-red cutoff is known to
arise in very reasonable extensions and refinement of the predictions
from simple inflationary scenarios, such as the modification to the
power spectrum from a pre-inflationary radiation dominated epoch or
from a sharp change in slope of the inflaton
potential~\cite{sin_sour04}.

\subsection{Gaussian primordial perturbations}
\label{gauss}

The detection of primordial non-Gaussian fluctuations in the CMB would
have a profound impact on our understanding of the physics of the
early universe. The Gaussianity of the CMB anisotropy on large angular
scales directly implies Gaussian primordial
perturbations~\cite{mun95,sper_gol99} that is theoretically motivated by
inflation~\cite{inflpert}. The simplest inflationary models predict
only very mild non-Gaussianity that should be undetectable in the
WMAP data.

The CMB anisotropy maps (including the non Gaussianity analysis
carried out by the WMAP team on the first year data~\cite{kom_wmap03})
have been found to be consistent with a Gaussian random field.
Consistent with the predictions of simple inflationary theories, no
significant deviations from Gaussianity in the CMB maps using general
tests such as Minkowski functionals, the bispectrum, trispectrum in
the three year WMAP data~\cite{sper_wmap06}.

\subsection{Primordial tensor (GW) perturbations}

Inflationary models can produce tensor perturbations from
gravitational waves that are predicted to evolve independently of the
scalar perturbations, with an uncorrelated power spectrum.  The
amplitude of a tensor mode falls off rapidly on sub-Hubble radius
scales. The tensor modes on the scales of Hubble-radius the line of
sight to the last scattering distort the photon propagation and
generate an additional anisotropy pattern predominantly on the largest
scales. It is common to parameterize the tensor component by the ratio
$r_{k_*} = A_{\rm t}/A_{\rm s}$, ratio of $A_{\rm t}$, the primordial
power in the transverse traceless part of the metric tensor
perturbations, and $A_{\rm s}$, the amplitude scalar perturbation at a
comoving wavenumber, $k_*$ (in $\mpc^{-1}$).  For power-law models,
recent WMAP data alone puts an improved upper limit on the tensor to
scalar ratio, \ensuremath{r_{0.002} < 0.55 \mbox{ } (95\%\mbox{\ CL})}
and the combination of WMAP and the lensing-normalized SDSS galaxy
survey implies \ensuremath{r_{0.002} < 0.28 \mbox{ } (95\%\mbox{\
CL})} ~\cite{boom_polar}.

\begin{figure}[h]
\begin{center}
\includegraphics[scale=0.4]{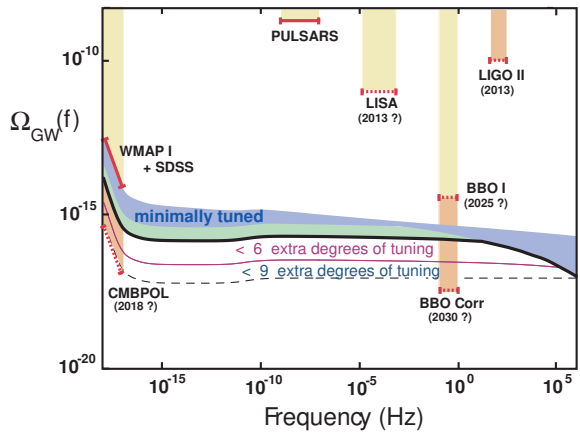} 
\caption{The figure taken from 
shows the theoretical predictions and observational constraints on
primordial gravitational waves from inflation. The gravitational wave
energy density per logarithmic frequency interval, (in units of the
critical density) is plotted versus frequency. The blue region
represents the range predicted for simple inflation models with the
minimal number of parameters and tunings. The dashed curves have lower
values of tensor contribution, $r$, that is possible with more fine
tuned inflationary scenarios.  The currently existing experimental
constraints shown are due to: big bang nucleosynthesis (BBN), binary
pulsars, and WMAP-1 (first year) with SDSS. Also shown are the
projections for LIGO (both LIGO-I, after one year running, and
LIGO-II); LISA; and BBO (both initial sensitivity, BBO-I, and after
cross-correlating receivers, BBO-Corr). Also seen the projected
sensitivity of a future space mission for CMB polarization (CMBPol).}
\label{SGWBspec} 
\end{center}
\end{figure}

On large angular scales, the curl component of CMB polarization is a
unique signature of tensor perturbations.  The CMB polarization is a
direct probe of the energy scale of early universe physics that
generate the primordial metric perturbations.  Inflation generates
both (scalar) density perturbations and (tensor) gravity wave
perturbations. The relative amplitude of tensor to scalar
perturbations, $r$, sets the energy scale for inflation $\EI =
3.4\times 10^{16}$~GeV~$r^{1/4}$.  A measurement of $B$--mode
polarization on large scales would give us this amplitude, and hence
{\em a direct determination of the energy scale of inflation.}
Besides being a generic prediction of inflation, the cosmological
gravity wave background from inflation would be a fundamental test of
GR on cosmic scales and the semi--classical behavior of gravity.
Figure~\ref{SGWBspec} summarizes the current theoretical
understanding, observational constraints and future possibilities for
the stochastic gravity wave background from Inflation.

\section{Gravitational instability mechanism for structure formation}
\label{GI}

It is a well accepted notion that the large scale structure in the
distribution of matter in the present universe arose due to
gravitational instability from the same primordial perturbation seen
in the CMB anisotropy at the epoch of recombination. This fundamental
assumption in our understanding of structure formation has recently
found an irrefutable direct observational
evidence~\cite{eis_sdss05,col_2Df05}.

For baryonic density comparable to that expected from Big Bang
nucleosynthesis, acoustic oscillations in the baryon-photon plasma
will also be observably imprinted onto the late-time power spectrum of
the non-relativistic matter.  
The remnants of the acoustic feature in the matter correlations are
weak ($10\%$ contrast in the power spectrum) and on large scales. The
acoustic oscillations of characteristic wavenumber translates to a
bump (a spike softened by gravitational clustering of baryon into the
well developed dark matter over-densities) in the correlation function
at $105 h^{-1}\mpc $ separation.  The large-scale correlation function
of a large spectroscopic sample of luminous, red galaxies (LRGs) from
the Sloan Digital Sky Survey that covers $\sim 4000$ square
degrees out to a redshift of $z\sim 0.5$ with $\sim 50,000$ galaxies
has allowed a clean detection of the acoustic bump in distribution of
matter in the present universe.  The acoustic signatures in the
large-scale clustering of galaxies provide direct, irrefutable
evidence for the theory of gravitational clustering, notably the idea
that large-scale fluctuations grow by linear perturbation theory from
$z\sim 1000$ to the present due to gravitational instability.

\section{Statistical Isotropy of the universe}
\label{SI}

The {\em Cosmological Principle} that led to the idealized FRW universe
found its strongest support in the discovery of the (nearly)
isotropic, Planckian, Cosmic Microwave Background. The isotropy around
every observer leads to spatially homogeneous cosmological models.
The large scale structure in the distribution of matter in the
universe (LSS) implies that the symmetries incorporated in FRW
cosmological models are to be interpreted statistically.


Interestingly enough, the statistical isotropy of CMB has come under a
lot of scrutiny after the WMAP results. Tantalizing evidence of SI
breakdown (albeit, in very different guises) has mounted in the {\it
WMAP} first year sky maps, using a variety of different statistics. It
was pointed out that the suppression of power in the quadrupole and
octopole are aligned \cite{maxwmap}.  Further ``multipole-vector''
directions associated with these multipoles (and some other low
multipoles as well) appear to be anomalously correlated
\cite{cop04,schw04}.  There are indications of asymmetry in the power
spectrum at low multipoles in opposite hemispheres
\cite{erik04a}. Possibly related, are the results of tests of
Gaussianity that show asymmetry in the amplitude of the measured genus
amplitude (at about $2$ to $3\sigma$ significance) between the North
and South galactic hemispheres \cite{erik04b,erik04c,par04}. Analysis
of the distribution of extrema in {\it WMAP} sky maps has indicated
non-gaussianity, and to some extent, violation of SI
\cite{lar_wan04}. The three-year WMAP maps are consistent with the
first-year maps up to a small quadrupole difference. The two
additional years of data and the improvements in analysis has not
significantly altered the low multipole structures in the
maps~\cite{hin_wmap06}. Hence, `anomalies' are expected to persist at
the same modest level of significance and are unlikely to be artifacts
of noise, systematics, or the analysis in the first year data.  The
cosmic significance of these `anomalies' remains debatable also
because of the aposteriori statistics employed to ferret them out of
the data. More importantly, what is missing is a common, well defined,
mathematical language to quantify SI (as distinct from non
Gaussianity) and the ability to ascribe statistical significance to
the anomalies unambiguously.

Recently, the Bipolar Power spectrum (BiPS) $\kappa_\ell$
($\ell=1,2,3, \ldots$) of the CMB map was proposed as a statistical
tool of detecting and measuring departure from
SI~\cite{us_apjl,us_pascos}. The non-zero value of the BiPS spectrum
imply the break down of statistical isotropy
\begin{equation} 
{\mathrm { STATISTICAL\,\,\,\, ISOTROPY}} \,\,\,\,\,\,\, \Longrightarrow \,\,\,\,\,\,\, 
\kappa_\ell\,=\,0 \,\,\,\,\,\,\, \forall \ell \ne 0.
\end{equation}
BiPS is sensitive to structures and patterns in the underlying total
two-point correlation function \cite{us_apjl,us_pascos,us_prl}. 


Measurement of the BiPS on the CMB anisotropy maps based the first
year WMAP data shows that the measured BiPS for all the WMAP sky maps
are consistent with statistical
isotropy~\cite{us_apjl2,us_apj,ghos06}. The ongoing BIPS analysis on
WMAP-3yr data indicates that BiPS of the three years maps show an
improvement in SI -- the deviations are smaller and
fewer~\cite{haj_sour06}.

CMB polarization maps over large areas of the sky have been recently
delivered by experiments in the near future. The statistical isotropy
of the CMB polarization maps will be an independent probe of the
cosmological principle.  Since CMB polarization is generated on at the
surface of last scattering, violations of statistical isotropy are
pristine cosmic signatures and more difficult to attribute to the local
universe.  The Bipolar Power spectrum has been defined and implemented
for CMB polarization and show great promise~\cite{bas06}.

\section{Conclusions}

The past few years has seen the emergence of a `concordant'
cosmological model that is consistent both with observational
constraints from the background evolution of the universe as well that
from the formation of large sale structures.  It is certainly fair to
say that the present edifice of the `standard' cosmological models is
robust. A set of foundation and pillars of cosmology have emerged and
are each supported by a number of distinct observations~\cite{me_jpo}.

The community is now looking beyond the estimation of parameters of a
working `standard' model of cosmology. There is increasing effort
towards establishing the basic principles and assumptions.  The
feasibility and promise of this ambitious goal is based on the grand
success in the recent years with the CMB anisotropy measurements. The
quest in the CMB sky from ground, balloon and space have indeed
yielded great results!  While the ongoing WMAP and up coming Planck
space missions will further improve the CMB polarization measurements,
there are already proposals for the next generation dedicated
satellite mission in 2015-20 for CMB polarization measurements at best
achievable sensitivity.

\section*{Acknowledgments}

I would like to thank the organizers for arranging an excellent
scientific meeting. It is a pleasure to thank and acknowledge the
contributions of students and collaborators.

\end{document}